\documentclass[rnote]{aa} % for the research notes
\usepackage{graphicx}
\usepackage{txfonts}
\begin{document}
   \title{Electron acceleration in a post-flare decimetric continuum source}

   \author{Prasad Subramanian,
          \inst{1}
          S. M. White,
          \inst{2}
M. Karlick\'{y},
\inst{3}
R. Sych,
\inst{4}
H. S. Sawant,
\inst{5}
S. Ananthakrishnan
\inst{6}
          }

   \offprints{Prasad Subramanian}

   \institute{Indian Institute of Astrophysics, Bangalore - 560034, India\\
              \email{psubrama@iiap.res.in}
         \and
             Dept. of Astronomy, University of Maryland, College Park, MD 20742, USA\\
             \email{white@astro.umd.edu}
\and
Astronomical Institute of the Academy of Sciences of the Czech Republic, 25165 Ond\u{r}ejov, Czech Republic \\
\email{karlicky@asu.cas.cz}
\and
Institute of Solar Terrestrial Physics, Siberian Branch of the Russian Academy of Sciences (ISTP), PO Box 4026, Irkutsk 33, 664033, Russia\\
\email{sych@iszf.irk.ru}
\and
Instituto Nacional de Pesquisas Espaciais, 515, 12201-970, San Jose Dos Campos, SP, Brazil\\
\email{sawant@das.inpe.br}
\and
National Centre for Radio Astrophysics, Tata Institute of Fundamental Research, Pune University Campus, PO Bag 3, Ganeshkhind, Pune 411007, India\\
\email{ananth@ncra.tifr.res.in}
}

   \date{}

% \abstract{}{}{}{}{} 
% 5 {} token are mandatory
 
  \abstract
   {}
   {To calculate the power budget for electron acceleration and the efficiency of the plasma emission mechanism in a post-flare decimetric continuum source.}
   {We have imaged a high brightness temperature ($\sim 10^{9}$K) post-flare source at 1060 MHz with the Giant Metrewave Radio Telescope (GMRT). We use information from these images and the dynamic spectrum from the Hiraiso spectrograph together with the theoretical method described in
Subramanian \& Becker (2006) to calculate the power input to the electron acceleration process. The
method assumes that the electrons are accelerated via a second-order Fermi acceleration mechanism.}
   {We find
  that the power input to the nonthermal electrons is
in the range $3\times 10^{25}$--$10^{26}$ erg/s. The efficiency of the overall plasma emission process starting from electron acceleration and culminating in the observed emission could range from $2.87\times 10^{-9}$ to $2.38 \times 10^{-8}$.}
   {}

\authorrunning{Subramanian et al.}
   \maketitle
%
%________________________________________________________________

\section{Introduction}
We observed the 1060 MHz counterpart of an M2.8 flare with the
Giant Metrewave Radio Telescope 
(GMRT) on Nov 17 2001. This long-duration flare started at 04:49 UT, peaked at 05:25 UT and ended at 06:11 UT.
This flare was accompanied by a fast halo coronal mass ejection (CME). 
The observations of this flare have been described in detail in Subramanian et al (2003). They studied the initiation of the flare-CME event and 
found evidence for the formation of a current sheet
below the erupting arcade that constitutes the CME. 
We concentrate our attention on a relatively simple morphology, high brightness temperature source that dominates the 1060 MHz emission after 05:30 UT. This source is part of a decimetric continuum feature that extends from $\sim$ 700-1100 MHz. The high brightness temperature of this source leads us to postulate a plasma emission mechanism for this source. We estimate the power input to the accelerated electrons that cause this emission and evaluate the overall efficiency of the process.

The plasma emission process is thought to be operative in several types of meter wavelength emission such as type 1/noise storm emission, type II and type III emission (e.g., Melrose 1975; Gary \& Hurford 2004). The overall emission process comprises of several steps: firstly, it is usually assumed that a distribution of nonthermal electrons exists. If these nonthermal electrons have a distribution that is anisotropic in velocity and/or physical space, they are capable of generating a high intensity population of Langmuir waves. These Langmuir waves, in turn, coalesce with a suitable population of low frequency waves to produce the observed radio emission. The efficiency of the Langmuir wave--observed radio emission part of this overall process has been well studied (e.g., Robinson, Cairns \& Willes 1994; Mitchell, Cairns \& Robinson 2003). In the context of the plasma emission process, the electron acceleration process that produces the nonthermal electron population in the first place has received attention relatively recently (Subramanian \& Becker 2004; 2006). Subramanian \& Becker (2004, 2006) have calculated the power input to the electron acceleration process and the efficiency of the overall process (starting from electron acceleration and culminating in the observed radiation) for noise storm continua. We use much more reliable observational inputs to calculate such an efficiency for the post-flare decimetric continuum we have observed. Accurate estimates of the efficiency of the overall plasma emission process are important, for instance, in deducing the shock strength from the intensity of observed type II emission. If the observed intensity can be well related to the energy in the nonthermal electrons, one can place reliable constraints on the parameters of the shock that produced the nonthermal electrons. Apart from the conventional type II emission, it is conceivable that meter wavelength emission from standing shocks in the vicinity of a reconnection region (Aurass et al. 2002) can also arise from the plasma emission process.

\section{Summary of radio observations}
Subramanian et al (2003) show that the 1060 MHz emission from this flare arises from two distinct sources in the GMRT images. The first two peaks of the 1060 MHz total lightcurve (top panel of figure 1) arise from the N-S oriented source shown in the first two panels of figure 9, Subramanian et al. (2003).
We concentrate here on the post-flare 1060 MHz emission (after $\sim$ 05:30 UT) that corresponds to the third broad peak of the total lightcurve (top panel of figure 1). The emission after 05:30 UT 
arises primarily from the southern (S) source shown in the last panel of figure 9 of Subramanian et al (2003). This is evident from the bottom panel of figure 1, where we have plotted the emission arising only from this S source.
The 1060 MHz contours of the S source are shown superposed on a soft X-ray SXI image in figure 2.

Figure 3 shows the dynamic spectrum corresponding to this event from the Hiraiso spectrograph. The 1060 MHz emission from the southern source imaged with the GMRT (figure 2) is part of the decimetric continuum shown in figure 3. The decimetric continuum emission spans a frequency range of approximately 700 to 1100 MHz.

\section{1060 MHz source size and brightness temperature}

We have fitted a two-dimensional Gaussian to the southern source shown in figure 2 and derived a full width at half-maximum (FWHM) size of $36^{''} \times 95^{''}$. The beam size
in the images is $30^{''}$, and we use this to get a deconvolved
source size of $20^{''} \times 90^{''}$. We use the 1 GHz Nobeyama Radio Polarimeter (NoRP)
lightcurve to determine that the peak post-flare flux (which, as
we know, arises almost exclusively from the S source) is
around 430 SFU. The brightness temperature $T_{b}$ (in K) of a source is
given by
\begin{equation}
T_{b} = 1.22 \times 10^{10}\,\frac{S}{f^{2}\,\theta\,\phi} \,\,\, ,
\label{eq1}
\end{equation} 
where $S$ is the flux in sfu, $f$ is the observing frequency in GHz, and $\theta$ and $\phi$ are the source dimensions in arcseconds. We use $f = 1.06$, $S = 430$, $\theta = 20$ and $\phi = 90$ to get a peak brightness temperature of $T_{b} = 2.7 \times 10^{9}$ K for the S source. The value for the peak flux $S$ is taken from the 1 GHz Nobeyama Radio Polarimeter (NoRP) data, since the GMRT data does not have absolute amplitude calibration. The high value of
the brightness temperature strongly suggests a plasma emission hypothesis for the observed radiation.

\section{Power input to electron acceleration process}
As mentioned earlier, it is commonly accepted that the starting point for the plasma emission process is a distribution of nonthermal electrons that is anisotropic in velocity space, and/or in physical space. An example of of the former is the ``gap'' distribution, which comprises of a thermal distribution together with a nonthermal
hump that is centered at a velocity $v_{c}$ that is considerably greater than $v_{e}$, the thermal velocity. The name derives from the
relatively empty part of the distribution between the thermal and the
nonthermal populations. Subramanian \& Becker (2006) have invoked a second-order Fermi acceleration process to account for the nonthermal electrons. They have self-consistently combined the requirements of the acceleration process with the requirements on the number of nonthermal electrons (relative to the number of thermal ones) to arrive at a 
upper limit for $v_c$. The acceleration timescale is lesser than the Coulomb loss timescale for electrons with velocities greater than $v_{c}$; these electrons therefore experience net acceleration, while those with velocities below $v_{c}$ experience (net) collisional losses and eventually join the thermal population. Subramanian \& Becker (2006) express their results 
in terms of the dimensionless energy $\xi_{c}$ that is related to $v_{c}$ in the following manner:
\begin{equation}
\xi_{c} = (1/2)\,m_{e}\,v_{c}^2/k\,T\,\,\, ,
\label{eq2}
\end{equation}
where the quantity $m_e$ is the electron mass, $k$ the Boltzmann constant and $T$ the coronal temperature. The upper limit on $\xi_{c}$ is denoted by $\xi_{\rm max}$, and is self-consistently 
determined as a function
of the spectral index $\alpha_{2}$ of the nonthermal electrons from the following equation (Subramanian \& Becker 2006), where the plasma frequency $\omega_{p}$ can be taken to be equal to the observing frequency for our purposes:

\begin{eqnarray}
&{4.4\,\omega_p \over T_e^{5/2}\,(\xi_{\rm max} + 1.5)} =
- {2 \, \xi_{\rm max}^{-\alpha_2/2} \, (3+\alpha_2) \,
\Gamma\left({3+\alpha_2 \over 2}
, \, \xi_{\rm max} \right) \over \sqrt{\pi} \ (3+2\alpha_2)}
\nonumber \\
&\phantom{lotsofspaaace} + \, 2  \, e^{-\xi_{\rm max}}
\left(\xi_{\rm max} \over \pi\right)^{1/2}
+ {\rm Erfc}\left(\xi_{\rm max}^{1/2}\right)
\, ,
\label{eq3}
\end{eqnarray}
where ${\rm Erfc}$ is the complementary error function. Under the plasma emission hypothesis, continuum emission at the fundamental will be polarized in the sense of the 'o' mode, while emission at the second harmonic will be polarized in the sense of the 'x' mode. Since observed type 1 continua are typically polarized in the sense of the 'o' mode, the continuum is assumed to be fundamental emission (Melrose 1980). We follow the same logic in assuming that the decimetric continuum radiation discussed in this paper is fundamental emission via the plasma emission process.
\subsection{Power input to nonthermal electrons $L_{\rm in}$}
 Once $\xi_{\rm max}$ is determined, the minimum nonthermal electron energy density $U_{*}$ (${\rm erg\,cm^{-3}}$) can be determined using the following equation:
\begin{eqnarray}
& {U_* \over n_e k_{\rm B} T_e} = \alpha_2 (3+\alpha_2)
\bigg[{2 \, \xi_{\rm max}^{(2-\alpha_2)/2} \, \Gamma\left({3+\alpha_2 \over 2}
, \, \xi_{\rm max} \right) \over \sqrt{\pi} \, (2-\alpha_2) (3+2\alpha_2)}
\nonumber \\
& \phantom{lotsofspaaace} + \, {2 \sqrt{\pi \xi_{\rm max}} \, (3 + 2 \, \xi_{\rm max})
\, e^{-\xi_{\rm max}} + \, 3 \pi \, {\rm Erfc}\left(\xi_{\rm max}^{1/2}\right)
\over 2 \pi (\alpha_2^2 + 3 \alpha_2 - 10)}\bigg] \ ,
\label{eq4}
\end{eqnarray}

 The minimum value of the power input $L_{\rm in}$ (${\rm erg\,s^{-1}}$) to the electron acceleration process is related to $U_{*}$ by (Subramanian \& Becker 2006)
\begin{equation}
L_{\rm in} = 8\,V\,D_{0}\,U_{*} \, ,
\label{eq5}
\end{equation}
where
\begin{equation}
D_{0} = \frac{1.2\,\Lambda\,n_{e}}{\xi_{\rm max}^{3/2}\,T_{e}^{3/2}} \, .
\label{eq6}
\end{equation}
The quantity $V$ is the volume of the acceleration region, and $D_{0}$ is a constant with units of inverse time, that is equal to 1/4 of the acceleration timescale (Subramanian, Becker \& Kazanas 1996). The quantity $\Lambda$ is the usual Coulomb logarithm, $n_{e}$ the ambient electron density and $T_{e}$ the electron temperature.

We assume the acceleration volume to be the same as that from which the observed emission emanates. We envisage this region to be a rectangular column. The dimensions of the face of this column would be equal to the observed source size, $20^{''} \times 90^{''}$, which is equivalent to $\sim$ 15467 km $\times$ 69600 km. Its depth can be calculated from the bandwidth of the emission as evident from the Hiraiso dynamic spectrum (Fig 3). The dynamic spectrum shows that the post-flare emission feature starting at $\sim$ 05:30 UT that we have imaged with the GMRT at 1060 MHz ranges from $\sim$ 700 to 1100 MHz. Using the height-density model developed in Aschwanden \& Benz (1995), we find that this frequency extent corresponds to a depth of $\sim 6289$ km for the emission column. The volume of the acceleration region is thus taken to be
\begin{equation}
V = 15467 \times 69600 \times 6289 = 6.77 \times 10^{12}\,\,\,\,{\rm km}^{3} \, \, \, .
\label{eq7}
\end{equation}

We use $\Lambda = 29$, $T_{e} = 10^{6}$K, $\omega_{p} = 2\pi \times 1060$ MHz, an ambient electron density corresponding to a plasma frequency of 1060 MHz, $n_{e} = 1.393 \times 10^{10}$ ${\rm cm}^{-3}$. Using these values in equations \ref{eq3} -- \ref{eq7}, we get
\begin{equation}
10^{26} \gtrsim L_{\rm in} \gtrsim 3 \times 10^{25}\,\,\,\,\,\,{\rm erg\,s^{-1}}\,
\label{eq8}
\end{equation}
for $-10 < \alpha_{2} < -5$. The power law index $\alpha_{2}$ is defined in such a manner that it has to be $< -5$ in order for the power in the accelerated electron population to remain finite (Subramanian \& Becker 2004; 2006). Save for values of $\alpha_{2}$ approaching the limiting value of -5, $L_{\rm in}$ is not a very sensitive function of $\alpha_{2}$. It may be noted that we have assumed the parameters relevant to the 1060 MHz layer to hold throughout the emission volume in calculating $L_{\rm in}$.

\section{Power in observed radiation $L_{\rm out}$}

We now turn our attention to estimating the power in the observed
decimetric continuum. As noted earlier, the 1060 Mhz post-flare radiation imaged with the GMRT is part of the decimetric continuum feature starting at around 05:30 UT and spanning a frequency range of $\sim$ 700 to 1100 MHz (figure 3). The peak intensity of this post-flare emission is $S_{\rm peak} = 430$ SFU (figure 1). We assume that this is the intensity of radiation emitted throughout the bandwidth of 1100 - 700 = 400 MHz. The radiation flux is therefore $F_{\rm out} = 400 \times 10^{6} \times 430$ SFU = $1.72 \times 10^{-11}$ ${\rm W\,m^{-2}}$. The power in the observed radiation $L_{\rm out}$ can be related to $F_{\rm out}$ using (e.g., Elgaroy 1977; Subramanian \& Becker 2006)
\begin{equation}
L_{\rm out} = F_{\rm out}\,R^{2}\,\Omega\,e^{-\tau} \, \, ,
\label{eq9}
\end{equation}
where $R$ is the sun-earth distance, $\Omega$ is the solid angle through which the radiation is beamed and $\tau$ is the optical depth of the source. Adopting a value of 0.6 steradian for $\Omega$ and 1 for $\tau$, we get $L_{\rm out} = 8.54 \times 10^{17}$ ${\rm erg\,s^{-1}}$.
The efficiency $\eta \equiv L_{\rm out}/L_{\rm in}$ of the overall process is thus
\begin{equation}
2.38 \times 10^{-8} \gtrsim \eta \gtrsim 2.87 \times 10^{-9}
\label{eq10}
\end{equation}

\section{Conclusions}
We have imaged a high brightness temperature decimetric continuum source at 1060 MHz with the GMRT. The emission from this source followed a M2.8 flare on Nov 17 2001. Its brightness temperature is $2.7 \times 10^{9}$ K and its observed dimensions are $20^{''} \times 90^{''}$. The emission is part of a decimetric continuum feature spanning a frequency range of $\sim$ 700-1100 MHz. The high brightness temperature suggests that this emission is due to nonthermal electrons emitting plasma waves that are subsequently converted into the observed electromagnetic radiation. We follow the formalism of Subramanian \& Becker (2006) in calculating the power input $L_{\rm in}$ to the second-order Fermi acceleration process that produces the nonthermal electrons. We use the source dimensions measured from the GMRT images and estimate the column depth of the acceleration region from the frequency extent of the decimetric continuum to find that $10^{26} \gtrsim L_{\rm in} \gtrsim 3 \times 10^{25}$ ${\rm erg\,s^{-1}}$. We find that the efficiency ($\eta$) of the overall process starting from electron acceleration and culminating in the observed radiation is in the range $2.38 \times 10^{-8} \gtrsim \eta \gtrsim 2.87 \times 10^{-9}$.

The numbers quoted here for $L_{\rm in}$ and $\eta$ are the best estimates of these quantities for the plasma emission process that we are aware of. There have been attempts to calculate these quantities for noise storm continua (e.g., Subramanian \& Becker 2006), but the observational inputs for these estimates were not as good as the ones used here. In particular, they used oft-quoted, but nonetheless rather rough estimates for the emission volume $V$ and the output power $L_{\rm out}$. These quantities are determined much more reliably from observations for the calculations presented here.

Before concluding, it is appropriate to reiterate the following caveats: one of the key observational inputs to this calculation is the size of the post-flare decimetric continuum source as determined from the 1060 MHz GMRT images. 
The $20^{''} \times 90^{''}$ source is
significantly more elongated in one dimension as compared to the other, suggesting that it might be oriented along a loop. The actual transverse dimension might then be somewhat smaller than the measured size of $20^{''}$. Scattering due to coronal turbulence is known to be anisotropic (e.g., Chandran \& Backer 2002). The contrast between the longitudinal and transverse dimensions might thus also be accentuated by preferential scattering along the longitudinal dimension. The other important observational input we use in our calculations is the column depth of the radiating source, which we calculate in an approximate manner from the extent of the decimetric continuum feature on the Hiraiso dynamic spectrograph (figure 3). In calculating the power in the observed radio emission $L_{\rm out}$, we also assume that the flux at 1 GHz is the same as that emitted throughout the $\sim$ 700--1100 MHz extent of the decimetric continuum emission. In the absence of calibrated multifrequency measurements, these are the most reasonable assumptions we can make. The formalism of Subramanian \& Becker (2006) that we have used to calculate the power fed to the nonthermal electrons, $L_{\rm in}$ accounts only for anisotropy in velocity space of the nonthermal electrons (the gap distribution). In reality, the nonthermal electrons are likely to be anisotropic both in physical space (as in a losscone distribution) as well as in velocity space (e.g., Thejappa 1991; Zaitsev et al. 1997). Finally, we have
asumed a coronal temperature of 1 MK in our calculations. There are indications that the temperature
might be as high as 2 MK, especially above active regions (e.g., Hayashi et al. 2006). As shown in Subramanian \& Becker (2006), doubling the value of the ambient coronal temperature will result in the efficiency estimate of the overall process being enhanced by a factor of $\sim$ 10.

%\begin{acknowledgements}
We appreciate insightful comments from the anonymous referee that has improved the content of
this paper.
%\end{acknowledgements}

\begin{figure*}[p]
\includegraphics[width=7.truein]{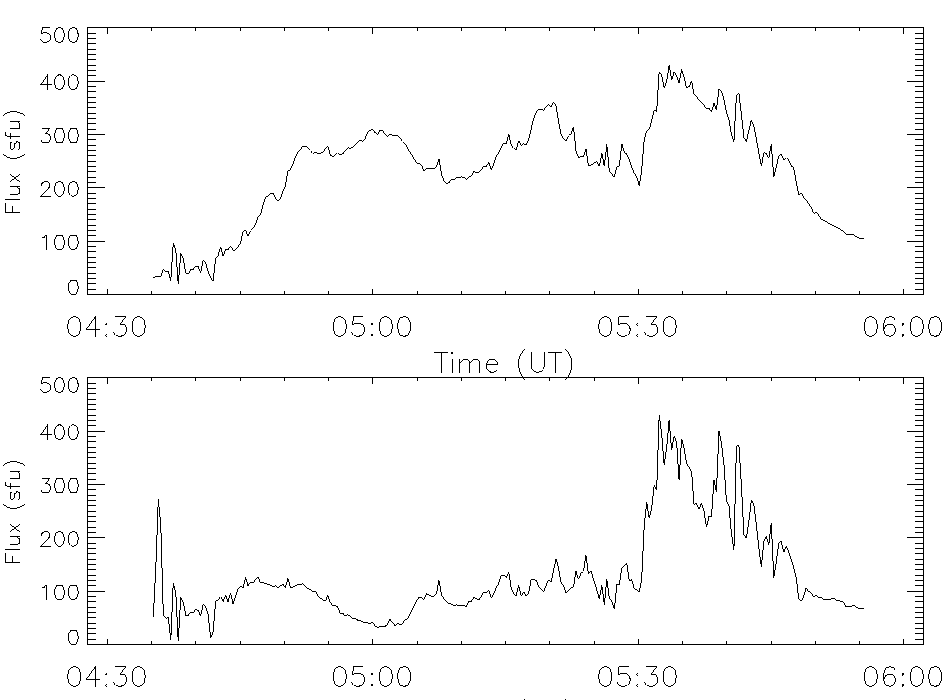}
\caption{1060 MHz lightcurves from the GMRT images for the M2.8 flare on Nov 17 2001. Since the GMRT images are not amplitude calibrated, we have normalized the amplitudes to the peak of the 1000 MHz lightcurve from the Nobeyama Radio Polarimeter (NoRP). {\em Top panel}: Total lightcurve for all the sources (figure 8, Subramanian et al. 2003). {\em Bottom panel}: Lightcurve for the southern (S) source only. It is evident that the high brightness temperature S source is the dominant contributor to the part of the lightcurve after $\sim$ 5:30 UT.}
\end{figure*}

\begin{center}
\begin{figure*}[p]
\includegraphics[height=4.truein]{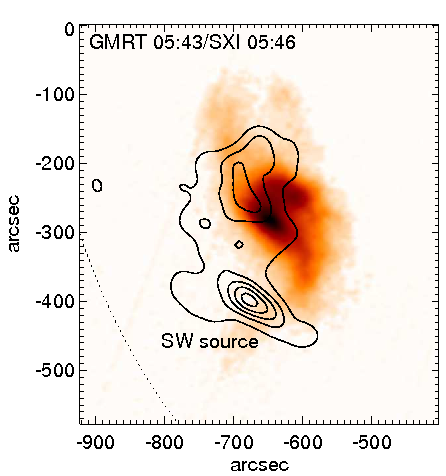}
\caption{Contours for the 1060 MHz southern (S) source overlaid on an SXI soft X-ray image.}
\end{figure*}
\end{center}

\begin{figure*}[p]
\includegraphics[width=7.truein]{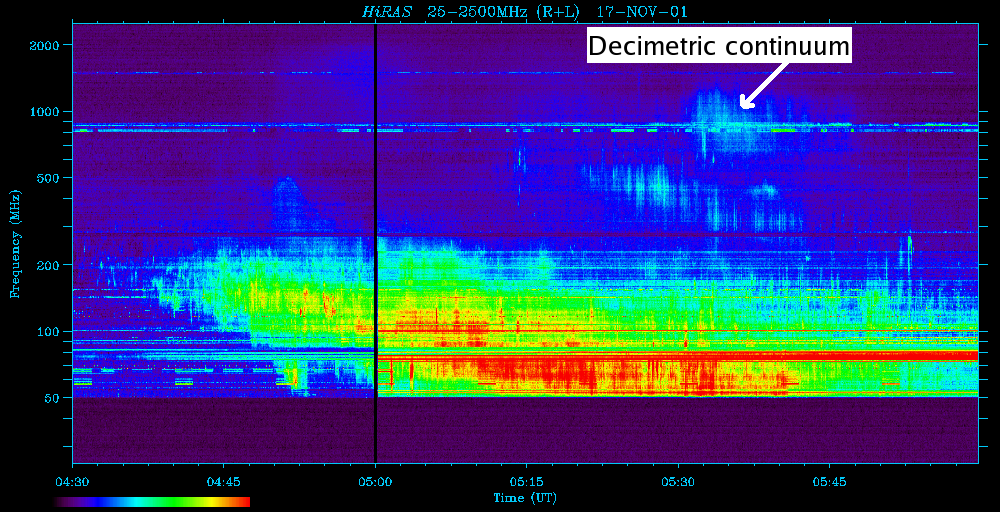}
\caption{The dynamic spectrum for this event from the Hiraiso spectrograph. The decimetric continuum emission feature is labelled in the figure.}
\end{figure*}

\end{document}